\documentclass[twocolumn,showpacs,preprintnumbers]{revtex4}
\usepackage{amssymb}
\usepackage{amsmath}
\usepackage{graphicx}
\usepackage{dcolumn}
\usepackage{bm}

\begin{document}

\title{Genuine tripartite entanglement and quantum phase transition}
\author{Chang-shui Yu}
\author{He-shan Song}
\email{hssong@dlut.edu.cn}
\author{Hai-tao Cui}
\affiliation{Department of Physics, Dalian University of Technology,\\
Dalian 116024, China}
\date{\today }

\begin{abstract}
A new formulation called as entanglement measure for simplification, is
presented to characterize genuine tripartite entanglement of $(2\times
2\times n)-$dimensional quantum pure states. The formulation shows that the
genuine tripartite entanglement can be described only on the basis of the
local $(2\times 2)-$dimensional reduced density matrix. In particular, the
two exactly solvable models of spin system studied by Yang (Phys. Rev. A
\textbf{71}, 030302(R) (2005)) is reconsidered by employing the entanglement
measure. The results show that a discontinuity in the first derivative of
the entanglement measure or in the entanglement measure itself of the ground
state just corresponds to the existence of quantum phase transition, which
is obviously prior to concurrence. Hence, the given entanglement measure may
become a new alternate candidate to help study the connection between
quantum entanglement and quantum phase transitions.
\end{abstract}

\pacs{03.67.Mn, 03.65.Ud, 05.70.Jk}
\maketitle

\bigskip

\begin{center}
\textbf{I. INTRODUCTION}
\end{center}

\medskip

Quantum entanglement is one of the most fascinating features of quantum
mechanics and a crucial physical resource in many quantum information
processing. It has attracted much attention in recent years. Even though a
great deal of effort has been made to characterize quantitatively the
entanglement properties of a quantum system [1-12], however, the good
understanding is still restricted in low-dimensional systems. The
quantification of entanglement for higher dimensional systems and
multipartite quantum systems remains to be an open question.

It has been shown that there are some good reasons to study entanglement in
multipartite quantum systems [13]. In particular, quantum entanglement in
the ground state of strongly correlated systems has attracted many
physicists' interest [14-25]. Most of the works focused on the spin models.
They employed the remarkable bipartite entanglement measure-\textit{%
concurrence} [1]-to measure the entanglement of a pair of nearest-neighbor
particles (in this paper, "concurrence" always refers to the concurrence of
a pair of nearest-neighbor particles for simplification, if there are not
other statements.) and attempted to understand the connection between
quantum entanglement (QE) and quantum phase transitions (QPT). Even though
it has been found in some works (for example, Refs. [22,25]) that the
critical behaviors of the concurrence such as a discontinuity in concurrence
or its first derivative of the ground state can signal a QPT (first-order
QPT (1QPT) or second-order QPT (2QPT) may be included), they are usually not
universal. A lot of examples [14,16-18,26] have shown that the critical
behaviors of concurrence of the ground state can not faithfully reflect the
QPT of the given models. In particular, in Ref. [18], the author found that
the discontinuity of the first derivative of concurrence for the exactly
solvable quantum spin-$\frac{1}{2}XY$ model with three-spin interactions
[27] does not signal any quantum critical points. They also showed that the
discontinuity of the first derivative of concurrence for the $XXZ$ spin
chain corresponds to a 1QPT instead of 2QPT. Hence, concurrence may not be a
good candidate to connect QE with QPT, even though concurrence indeed does
well in some models. In fact, QPT should be an embodiment of some collective
behaviors of a multipartite systems, while concurrence only describes some
relation (entanglement or separability) between a pair of local particles.
It is necessarily a shortcoming for concurrence to capture some collective
behaviors [13]. It is naturally expected that some other entanglement
measure can be presented to help reveal the connection with QPTs.

In this paper, we present a new formulation to characterize the entanglement
of tripartite $(2\times 2\times n)-$dimensional quantum pure states. It has
been proved to be an equivalent expression to that in Ref. [28]. That is to
say, not only can the formulation here characterize the properties of
genuine tripartite entanglement, but also it can be considered as a "good"
entanglement measure if one does not consider local operations in the
higer-dimensional subsystem (local unitary transformations excluded).
However, the distinct advantage of the current formulation is that it can
significantly simplify that in Ref. [28]. To measure the genuine tripartite
entanglement, it is not necessary to obtain the total density matrix of the $%
(2\times 2\times n)-$dimensional quantum pure states, but only the local $%
(2\times 2)$-dimensional reduced density matrix. As applications, we
reconsider the two models of spin system presented in Ref. [18] and employ
our measure to calculate the genuine tripartite entanglement by considering
the whole spin chain as a $(2\times 2\times n)-$dimensional quantum pure
state. The results indicate that our measure can faithfully signal the 2QPT
point of the spin-$\frac{1}{2}XY$ model and the 1QPT point at $\Delta =-1$
of the $XXZ$ model. In this sense, our measure is better than concurrence
and becomes a new alternate entanglement measure to study the connection
between QE and QPT. The paper is organized as followed. We first introduce
the new formulation for genuine tripartite entanglement and then apply\ the
measure to the two spin models; The conclusion is drawn in the end.

\bigskip

\begin{center}
\textbf{II. THE NEW FORMULATION OF GENUINE TRIPARTITE ENTANGLEMENT MEASURE}
\end{center}

\medskip

\bigskip Consider a tripartite $\left( 2\times 2\times n\right) -$%
dimensional pure state given in computational basis by
\begin{equation}
\left\vert \psi _{ABC}\right\rangle
=\sum_{i,j=0}^{1}\sum_{k=0}^{n-1}a_{ijk}\left\vert i\right\rangle
_{A}\left\vert j\right\rangle _{B}\left\vert k\right\rangle _{C},
\end{equation}%
with $\sum_{i,j=0}^{1}\sum_{k=0}^{n-1}\left\vert a_{ijk}\right\vert ^{2}=1$,
the reduced density matrix by tracing over party $C$ is denoted by $\rho
_{AB}$ which is a $\left( 2\times 2\right) -$dimensional matrix. Based on
the Pauli matrix $\sigma _{y}=\left(
\begin{array}{cc}
0 & -i \\
i & 0%
\end{array}%
\right) $, the spin-flipped state denoted by $\tilde{\rho}_{AB}$ can be
obtained by%
\begin{equation}
\tilde{\rho}_{AB}=\left( \sigma _{y}\otimes \sigma _{y}\right) \rho
_{AB}^{\ast }\left( \sigma _{y}\otimes \sigma _{y}\right) ,
\end{equation}%
where $\rho _{AB}^{\ast }$ stands for the complex conjugate of $\rho _{AB}$.

\textbf{Theorem 1.} \textit{The genuine tripartite entanglement }$\tau
(\left\vert \psi _{ABC}\right\rangle )$\textit{\ of }$\left\vert \psi
_{ABC}\right\rangle $\textit{\ defined in }$\left( 2\times 2\times n\right) -
$\textit{dimensional Hilbert space} \textit{can be characterized by }%
\begin{equation}
\tau (\left\vert \psi _{ABC}\right\rangle )=\sqrt[4]{\left[ tr\left( \rho
_{AB}\tilde{\rho}_{AB}\right) \right] ^{2}-tr\left[ \left( \rho _{AB}\tilde{%
\rho}_{AB}\right) ^{2}\right] },
\end{equation}%
\textit{where }$tr\left( \cdot \right) $\textit{\ denotes the trace
operation of a matrix.}

\textbf{Proof.} For a pure state $\left\vert \psi _{ABC}\right\rangle $, the
corresponding reduced density matrix $\rho _{AB}$ can usually be expressed
as a bipartite mixed state as
\begin{equation}
\rho _{AB}=\sum_{i}p_{i}\left\vert \varphi _{i}\right\rangle \left\langle
\varphi _{i}\right\vert ,\sum_{i}p_{i}=1,
\end{equation}%
according to any decomposition of $\rho _{AB}$ (Note that $\rho _{AB}$ may
be a pure state, which can be considered as a special case of eq. (4) with $%
p_{1}=1$ and $p_{i\neq 1}=0$). The matrix notation of eq. (4) can be given
by $\rho _{AB}=\Psi W\Psi ^{\dag }$, where the columns of $\Psi $
corresponds to $\left\vert \varphi _{i}\right\rangle $ and $W$ is a diagonal
matrix with its diagonal entries corresponding to $p_{i}$ respectively.
Therefore, eq. (5) can be rewritten by
\begin{eqnarray}
\tau ^{4}(\left\vert \psi _{ABC}\right\rangle ) &=&\left\{ tr\left[ \Psi
W\Psi ^{\dag }\left( \sigma _{y}\otimes \sigma _{y}\right) \Psi ^{\ast
}W\Psi ^{T}\left( \sigma _{y}\otimes \sigma _{y}\right) \right] \right\} ^{2}
\notag \\
&-&tr\left[ \Psi W\Psi ^{\dag }\left( \sigma _{y}\otimes \sigma _{y}\right)
\Psi ^{\ast }W\Psi ^{T}\left( \sigma _{y}\otimes \sigma _{y}\right) \right]
^{2},
\end{eqnarray}%
with the superscript $T$ denoting transpose operation. By a small change of
eq. (3), one can obtain
\begin{widetext}
\begin{eqnarray}
\tau ^{4}(\left\vert \psi _{ABC}\right\rangle ) &=&\left\{ tr\left[
\sqrt{W}\Psi ^{\dag }\left( \sigma _{y}\otimes \sigma _{y}\right)
\Psi ^{\ast }\sqrt{W}\cdot \sqrt{W}\Psi ^{T}\left( \sigma
_{y}\otimes \sigma _{y}\right) \Psi \sqrt{W}\right] \right\} ^{2} \notag \\
&-&tr\left[ \sqrt{W}\Psi ^{\dag }\left( \sigma _{y}\otimes \sigma
_{y}\right) \Psi ^{\ast }\sqrt{W}\cdot \sqrt{W}\Psi ^{T}\left(
\sigma _{y}\otimes \sigma _{y}\right) \Psi \sqrt{W}\right] ^{2},
\end{eqnarray}%
\end{widetext}where $\sqrt{W}$ stands for the squared root of the entries of
$W$.

Recalling the procedure of constructing the genuine tripartite entanglement
in Ref. [28], one has to project $\left\vert \psi _{ABC}\right\rangle $ onto
the subspace of Party $C$ and obtain a set of unnormalized bipartite pure
states. If assuming every element of the set just corresponds to $\sqrt{p_{i}%
}\left\vert \varphi _{i}\right\rangle $ given by eq. (4), in other words, $%
\Psi \sqrt{W}$ is just the matrix notation of $\{\sqrt{p_{i}}\left\vert
\varphi _{i}\right\rangle \}$, one can easily find that
\begin{equation}
\tau ^{4}(\left\vert \psi _{ABC}\right\rangle )=\left[ tr\left( \mathcal{MM}%
^{\dag }\right) \right] ^{2}-tr\left[ \left( \mathcal{MM}^{\dag }\right) ^{2}%
\right] ,
\end{equation}%
where $\mathcal{M=}\sqrt{W}\Psi ^{T}\left( \sigma _{y}\otimes \sigma
_{y}\right) \Psi \sqrt{W}$. Eq. (7) is consistent to that in Ref. [28].
Therefore, $\tau (\left\vert \psi _{ABC}\right\rangle )$ is an entanglement
semi-monotone of genuine tripartite entanglement.$\hfill {}\Box $

The distinct advantage of current version given in eq. (3) is that $\tau
(\left\vert \psi _{ABC}\right\rangle )$ can be easily obtained from the
reduced density matrix instead of obtaining the total pure state $\left\vert
\psi _{ABC}\right\rangle $ and following the complicated procedure given in
Ref. [28]. In fact, eq. (3) is also a simplification of the genuine
tripartite entanglement measure of $\left( 2\times 2\times 2\right) -$%
dimensional pure states given in Ref. [5]. The advantage is something like
the case of bipartite entanglement measure, i.e. the linear entropy (or its
analogue [12]) of a bipartite entangled pure state can be considered as an
analogous simplification\ [1] of the length of concurrence vector [3]. One
will find that the simplification makes the application of our measure more
convenient. Furthermore, for the extension of eq. (3) to mixed states, one
will have to turn to the same procedure given in Ref. [28].

\bigskip

\begin{center}
\textbf{III. THE SIGNAL OF QUANTUM PHASE TRANSITION}
\end{center}

\medskip

Now, let us reconsider the isotropic spin-$\frac{1}{2}XY$ chain with
three-spin interaction presented in Refs. [18, 27], which is an exactly
solvable quantum spin model. The Hamiltonian is
\begin{equation}
H=-\sum_{i=1}^{N}\left[ \sigma _{i}^{x}\sigma _{i+1}^{x}+\sigma
_{i}^{y}\sigma _{i+1}^{y}+\frac{\lambda }{2}\left( \sigma _{i-1}^{x}\sigma
_{i}^{z}\sigma _{i+1}^{y}-\sigma _{i-1}^{y}\sigma _{i}^{z}\sigma
_{i+1}^{x}\right) \right] ,
\end{equation}%
where $N$ is the number of sites, $\sigma _{i}^{\alpha }$($\alpha =x,y,z$)
are the Pauli matrices, and $\lambda $ is a dimensionless parameter
characterizing the three-spin interaction strength. Here the periodic
boundary condition $\sigma _{N+1}=\sigma _{1}$ is assumed. Ref. [27] has
shown that the three-spin interaction can lead to a 2QPT at $\lambda
=\lambda _{C}=1$. However, Ref. [18] has shown that the discontinuity of the
first derivative of the ground-state concurrence of the nearest--neighbor
spins can not dependably signal a QPT, because there does not exist any QPT
at $\lambda =2/(\sqrt{2}-1)\pi $, but the first derivative of the
ground-state concurrence shows discontinuity here. Furthermore, Ref. [18]
has shown that the von Neumann entropy [23] as an entanglement measure
defined as $S=-tr(\rho _{j}\log _{2}\rho _{j})$ also fails to detect the QPT
of the current model, where $\rho _{j}$ is the one-particle reduced density
matrix. It is natural to wonder whether the entanglement measure given by
eq. (3) can well detect the QPT of the model.

Since one can consider the total ground state as a tripartite $\left(
2\times 2\times n\right) -$dimensional pure state, in particular that for a
given spin system it is not necessary to introduce local operations to such
as Positive Operator Values Measure as so on, it will be convenient to
employ the entanglement measure presented above to measure the entanglement
and also investigate the QPT in the model. Here we consider such a grouping
as \textit{two-nearest-neighbor-particle} versus \textit{others}. Due to the
entanglement measure given by eq. (3), it is necessary to only obtain the
two-nearest-neighbor-particle reduced density matrix which has been in fact
given in Ref. [18]. On the basis of the Hamiltonian given in Ref. [18], the
two-nearest-neighbor-particle reduced density matrix by tracing over other
spins can be given by [21]%
\begin{equation}
\rho _{i,i+1}=\left(
\begin{array}{cccc}
u_{i,i+1} & 0 & 0 & 0 \\
0 & w_{i,i+1} & z_{i,i+1} & 0 \\
0 & z_{i,i+1} & w_{i,i+1} & 0 \\
0 & 0 & 0 & u_{i,i+1}%
\end{array}%
\right)
\end{equation}%
in the standard basis \{$\left\vert \uparrow \uparrow \right\rangle
,\left\vert \uparrow \downarrow \right\rangle ,\left\vert \downarrow
\uparrow \right\rangle ,\left\vert \downarrow \downarrow \right\rangle $\},
where
\begin{equation*}
u_{i,i+1}=\frac{1}{4}(1-G^{2}),
\end{equation*}%
\begin{equation*}
w_{i,i+1}=\frac{1}{4}(1+G^{2}),
\end{equation*}%
and%
\begin{equation}
z_{i,i+1}=G/2,
\end{equation}%
with
\begin{equation}
G=\left\{
\begin{array}{cc}
\frac{2}{\pi }, & \lambda <1, \\
\frac{2}{\pi \lambda }, & \lambda \geq 1.%
\end{array}%
\right.
\end{equation}%
Hence, the entanglement measure $\tau _{XY}$ can be easily obtained in terms
of eq. (3), which is given in Fig.1. Fig.1 shows that $\tau _{XY}$ is
continuous at all $\lambda $ and the first derivative of $\tau _{XY}$ is
only discontinuous at $\lambda =\lambda _{C}=1$, which indicates the 2QPT
only at $\lambda _{C}$.

\begin{figure}[tbp]
\includegraphics[width=7.5cm]{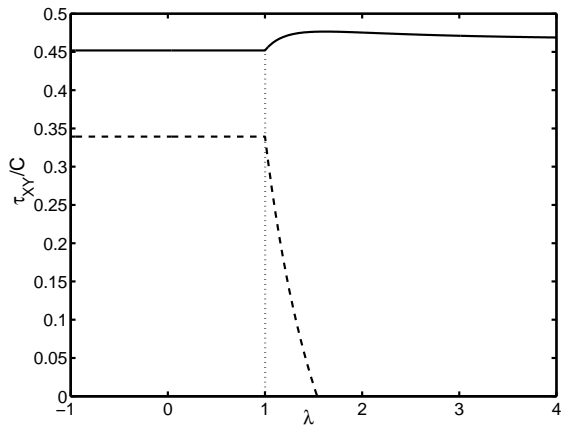}% Here is how to import EPS art
\caption{(Dimensionless) The genuine tripartite entanglement $\protect\tau%
_{XY}$ (solid line) and the concurrence $C$ (dashed line) of the ground
state of the isotropic spin-$\frac{1}{2}XY$ chain versus the three-spin
interaction strength $\protect\lambda$. The figure shows obviously that $%
\frac{\partial\protect\tau_{XY}}{\partial\protect\lambda}$ is discontinuous
at $\protect\lambda=1$, which signals only one 2QPT.}
\end{figure}

Another model considered in Ref. [18] is the one-dimensional $XXZ$ model,%
\begin{equation}
H_{XXZ}=\sum_{i=1}^{N}\left[ \sigma _{i}^{x}\sigma _{i+1}^{x}+\sigma
_{i}^{y}\sigma _{i+1}^{y}+\Delta \sigma _{i}^{z}\sigma _{i+1}^{z}\right] .
\end{equation}%
The model has been shown that there exist a 1QPT instead of 2QPT at the
critical point $\Delta =-1$. However, the concurrence of the
nearest-neighbor-two-particle reduced density matrix is continuous at $%
\Delta =-1$, while the first derivative of the concurrence is not continuous
at the critical point. The behavior of concurrence associated with its
derivative shows a 2QPT of the model, which is opposite to the fact, in this
sense, Ref. [18] concluded that the nonanalyticity is misleading for
detection of a QPT. However, one will find that the nonanalyticity of our
measure $\tau $ can just show a 1QPT of the model at $\Delta =-1$.

Based on Refs. [18, 21], the nearest-neighbor-two-particle reduced density
matrix has the same form to eq. (9). The elements can be obtained from Ref.
[18] as%
\begin{equation}
u_{i,i+1}=\frac{1}{4}(1+\frac{\partial \mathcal{E}}{\partial \Delta }),
\end{equation}%
\begin{equation}
z_{i,i+1}=\frac{1}{4}(\mathcal{E}-\Delta \frac{\partial \mathcal{E}}{%
\partial \Delta }),
\end{equation}%
where $\mathcal{E}=\left\langle \sigma _{i}^{x}\sigma
_{i+1}^{x}\right\rangle +$ $\left\langle \sigma _{i}^{y}\sigma
_{i+1}^{y}\right\rangle $ +$\Delta \left\langle \sigma _{i}^{z}\sigma
_{i+1}^{z}\right\rangle $ is the ground-state energy per site for the $XXZ$
model. At $\Delta =-1$, one has $\left. \mathcal{E}\right\vert _{\Delta
=-1}=-1$ [18, 29]. Substituting eq. (9) associated with eqs. (13, 14) into
eq. (3), one will obtain the corresponding genuine tripartite entanglement
measure $\tau _{XXZ}$ which can be formally written by
\begin{equation}
\tau _{XXZ}=f(\frac{\partial \mathcal{E}}{\partial \Delta }),
\end{equation}%
where $f(x)$ is a continuous function on $x$, which can be found by eq. (3).
From Refs. [18, 29], one will also find that $\left. \frac{\partial \mathcal{%
E}}{\partial \Delta }\right\vert _{\Delta \rightarrow -1^{+}}\rightarrow 0$
and $\left. \frac{\partial \mathcal{E}}{\partial \Delta }\right\vert
_{\Delta <-1}=1$. That is to say, $\frac{\partial \mathcal{E}}{\partial
\Delta }$ is not continuous at $\Delta =-1$. Thus one can conclude that $%
\tau _{XXZ}$ is not continuous at $\Delta =-1$ due to the property of $f(x)$%
. This result indicates that the discontinuity of $\tau _{XXZ}$ at $\Delta
=-1$ signals a 1QPT. However, the other critical point of the model at $%
\Delta =1$ does not correspond to the discontinuous behavior of $\tau _{XXZ}$
or the its first derivative, which is analogous to the concurrence. In fact,
numerical results based on Ref. [30,31] can show that $\tau _{XXZ}$ reaches
its minimum at $\Delta =1$. The result is not given here.

\bigskip

\begin{center}
\textbf{IV. CONCLUSION AND DISCUSSIONS}
\end{center}

\medskip

We have presented a new formulation to characterize genuine tripartite
entanglement of $(2\times 2\times n)-$dimensional quantum pure states. It
has been proved that the formulation is an equivalent description of the
genuine tripartite entanglement introduced in Ref. [28]. The distinct
advantage is that the current formulation can be obtained only by the local $%
(2\times 2)-$dimensional reduced density matrix, which significantly
simplifies the calculation of the genuine tripartite entanglement introduced
in Ref. [28] including the original one [5]. This is something like the case
of bipartite entanglement measure, i.e. the linear entropy as a
simplification of the length of concurrence vector can be obtained by the
reduced matrix. By employing the new measure, we reconsider the two exactly
solvable spin models presented in Ref. [18]. The results have shown that a
discontinuity in the first derivative of the entanglement measure or in the
entanglement measure itself of the ground state just corresponds to the
existence of quantum phase transitions, which is obviously prior to
concurrence as well as the von Neumann entropy. In this sense, the
entanglement measure may become a new alternate candidate to help study the
connection between quantum entanglement and quantum phase transitions. Of
course, the measure can not always do well, for example, at the critical
point $\Delta =1$ for $XXZ$ model, neither $\tau _{XXZ}$ nor $\frac{\partial
\tau _{XXZ}}{\partial \Delta }$ shows discontinuous behavior. From
block-block-block entanglement point of view, it is implied that the spin
chain is considered as tripartite quantum states including two $2$%
-dimensional blocks. The other treatments such as three higher-dimensional
blocks, even multiple higher-dimensional blocks, may be more effective for
the detection of QPT. In this sense, the generalization of $\tau $ to
higher-dimensional systems and multipartite systems seem to be needed.

\bigskip

\begin{center}
\textbf{V. ACKNOWLEDGEMENT}
\end{center}

\medskip

I would like to thank C. Li for his help. This work was supported by the
National Natural Science Foundation of China, under Grant Nos. 10575017 and
60472017.

\end{document}